\documentstyle [12pt] {article}

\begin{document}

\title{\large \bf 2D BLACK HOLES AND 2D GRAVITY
\footnotemark[1]}
\footnotetext[1]{Talk given at the Workshop on Low Dimensional Topology
and Quantum Fild Theory at the Isaac Newton Institute for Mathematical
Science in September 1992.}

\author{ Farhad Ardalan }
\date{}
\maketitle

\newpage
\begin {abstract}
{\small The SL$(2,R)/U(1)$ coset model, with $U(1)$
an element of
the third conjugacy class of $SL(2,R)$ subgroups, is considered.
The resulting theory is seen to
collapse to a one dimensional field theory of Liouville. Then the
2 dimensional
black hole
$SL(2,R)/U(1)$, with $U(1)$ a non-compact subgroup boosted by a
Lorentz transformation,
is considered. In the limit of high boost, the resulting black hole
is found to tend to
the Liouville field coupled to $a\ C=1$ matter field.
The limit of the vertex
operators of the 2 dimensional black hole also tend to those of the
$C=1$ two dimensional gravity.}
\end{abstract}
\newpage

The two dimensional black hole of the $SL(2,R)/U(1)$ coset model
has been obtained
by gauging a noncompact $U(1)$ subgroup of the $SL(2,R)WZW$
theory $[1]$ .
Gauging a compact subgroup yields a coset model which results
in the same
black hole solution when it is properly analytically continued.
These two
subgroups represent two of the three classes of conjugacies of
subgroups of $SL(2,R).$\\

It is therefore intriguing to study the result of gauging a
subgroup in the remaining
third conjuacy class. The situation is of course reminicent of the
three classes
of isotropy subgroups of the four dimensional Lorentz group,
for massive, tachyonic, and massless states. The third conjugacy
class corresponds to the massless class.
It will be found that the $L(2,R)WZW$ model gauged by this third
class of subgroups
results in a one dimensional theory which is a Liouville field
theory.\\

To understand this reduction of the two dimensional space to the
one dimensional
Liouville field theory, a $U(1)$ subgroup of $SL(2,R)$ is gauged
which is
boosted by a Lorentz transformation, with boost parameter t.
It is found that
as $t \longrightarrow \infty$, the $2d$ black hole in fact
collapses to one
dimensional Liouville; the collapse being the consequence of a
sudden enlargement
of the orbit of the gauge group.\\

For large values of t, the remaining degree of freedom
resembles $a\ C=1$
matter conformal field theory coupled to the Louville. Therefore in
the
large $t$ limit a concrete connection between the $2d$ black hole
theory and the $C=1$ matter
coupled to $2d$ gravity is established. The connection is utilized
to study the
relation between the vertex operators of the two theories;
in particular the
amplitudes for the scattering of strings in the two backgrounds
are related. It
is hoped that this connection could shed some light on the question
of the
exact relation between the spectrum of the two theories $[2,3,4]$.\\

To begin with, let us take $\sigma _1, i\sigma _2$ and $\sigma _3$
as the generators of $SL(2,R)$. The $i\sigma _2$ generates the
compact $U(1)$
subgroup which upon gauging gives the Euclidean black hole; and
$\sigma _3$
generates the noncompact subgroup whose gauging is responsible for
the Lorentzian
black hole $[1]$. An element of the third conjugacy class of $U(1)$
subgroups
is generated by $\sigma =\sigma _3+i\sigma _2$, which upon gauging
in the usual
manner [1], yields the following action:\\
\begin{equation}
\begin{array}{ll}
I(g,A)=I_{WZW}+\frac {k}{2\pi }\int dz^2 tr
(\overline {A} g^{-1}\partial g+
A\overline {\partial} g g^{-1}+g A g^{-1}\overline {A}), &  \\
I_{WZW}=
\frac {k}{4\pi }\displaystyle \int _{\sum } dz^2 tr (g^{-1}
\partial g g^{-1} \overline {\partial } g)-
\frac {k}{12\pi } \displaystyle \int _{B} d^2_ztr (g^{-1}dg\wedge
g^{-1}dg\wedge g^{-1}dg). & \end{array}
\end{equation}\\
where $g$ is an element of $SL(2,R)$ and $A$ is the gauge field
lying in the Lie
algebra generated by $\sigma $. Note that the usual $A\overline {A}$
term does not appear here because the element $\sigma $ is nilpotent.
It is
easily verified that this action is invariant under the axial
vector gauge
transformation,\\
\begin{equation}
\begin{array}{ll}
g\longrightarrow hgh, & A\longrightarrow h^{-1} (A+\partial )h,\\
 & \overline {A}\longrightarrow h^{-1} (\overline {A}+\overline
{\partial})h.\end{array}
\end{equation}
Now in the parametrization\\
\begin{equation}
g=\left (\begin{array}{ll}
a & u\\
-v & b \end{array}\right )
, uv+ab=1,
\end{equation}
of $g$, the gauge freedom (2) allows for a single gauge fixing
$a+b=0$, upon
whose imposition, and upon integration of the action over the
gauge fields
$A {\rm and} \overline {A}$, leads to the effective action,\\
\begin{equation}
I_{eff}=\frac {k}{8\pi }\int d^2z(\partial \varphi \overline
{\partial} \varphi +\alpha \varphi),
\end{equation}
which is that of a Liouville field. $\alpha $ is a constant which
is left undetermined due to
ambiguities in the determinant in the path integral.
Here $e^{\varphi }=a-b-u-v$. A striking feature of this action
is the absence of
an additional field expected on the basis of a simple counting
of the degrees
of freedom.\\

The Liouville field theory has been considered from the point of
view of $WZW$
gauging in previous works $[5,6]$, However, the gauging used in
these attempts
are of the from $g\longrightarrow h_L gh_R$, where $h_L$ and $h_R$
are independent and
correspond to positive and negative roots of $SL(2)$, respectively.
Therefore
it is natural that two degree of freedom are eliminated by gauging
the original
$SL(2,R)$ theory. This is to be contrasted with the gauging above
which is expected
to eliminate only one degree of freedom.\\

To understand this reduction of the degrees of freedom, let us
consider the
$2d$ black hole anew; this time with a $U(1)$ subgroup which is
a Lorentz
transformation of the original noncompact $U(1)$ subgroup,
$i.e.$ the group
generated by $\sigma _3^t$, where
\begin{equation}
\begin{array}{ll}
\sigma _3^t &=e^{-\frac {t}{2} \sigma _1}\sigma _3 e^{\frac {t}{2}
\sigma _1}\\
 & =ch t \sigma _3+sh t i\sigma _2.\end{array}
\end{equation}

As $t\longrightarrow \infty ,\ \sigma _3^t$ becomes proportional to
$\sigma $
defined above. Therfore it is expected that as $t\longrightarrow
\infty$,
the $2d$ black hole obtained from gauging the $U(1)$ generated by
$\sigma _3^t$,
should reduce to the Liouville theory as obtained above; moreover
the asymptotic behavior
should explain the reduction of the degees of freedom.\\
Note that, for finite $t$ this gauged $SL(2,R)$ theory, is
equivalent to the
original black hole theory with $t=0$, because subgroups with
different $t$
are in the same conjugacy class and the resultant coset theories
should
be equivalent.\\

To describe the black hole geometry of the gauged theory, a gauge
fixing is
requied. Recall that for the $U(1)$ gauging prior to boosting,
two regions of
the $(u,v)$ plane are distinguished by two distinct gauge conditions.
The first
region, consisting of the region $I-IV$ of $Ref.1$ is determined
by the gauge
condition $a=b$, in the parameterization (3); and the second region,
consisting of
the region $V$ and $VI$ of the above reference is determined by the
gauge condition
$a+b=0.$ Similarly for the boosted theory there are two regions,
the first one is
determined by the gauge condition $a-b=(u+v)$ th$t$; and the second
one isdetermined by the condition $a+b=0$. Now, using the $U(1)$
generator $(5)$
and integrating the gauge fields leads to a black hole geometry in
which the
horizon which originally lied on the $u$ and $v$ axes will now
tend towards each
other as $t$ gets larger and the singularity moves away from the
origin of
$(u,v)$ plane. The effective action in the large $t$ limit for the
gauge condition
$a+b=0$, will then read,\\
\begin{equation}
\begin{array}{ll}
I_{eff}=+\frac{k}{8\pi }\int dz^2 [\partial \varphi \overline
{\partial} \varphi +\alpha
\varphi -4e^{-2t} \partial x \overline
{\partial} x e^{-2 \varphi }-4e^{-2t}
(x^2+4)e^{-2\varphi }&  \\
(\partial \varphi \overline {\partial} \varphi +\beta )],&
\end{array}
\end{equation}
where again $\alpha $ and $\beta $ as constants to be determined
$e.g.$ a la
$DDK [8]$. The terms decreasing faster than $e^{-2t}$ have been
dropped in (6);
and $x=u-v$.\\

As $t\longrightarrow \infty $, only the first two terms in (6)
survive and the action becomes
identical to the Liouville action (4) obtained as a result of
gauging $\sigma $.
Now it is possible to trace the cause of the elimination of the
extra degree of
freedom $x$, when $t\longrightarrow \infty$: As it was pointed out
before,
when $t$ is finite there are two distinct regions of $(u,v)$
plane determined by gauges
$a-b=(u+v)$ th$t$ and $a+b=0$. When $t$ increases the region
described by
the second gauge fixing condition, region $V$ of Ref.1, narrows
and tends to a one
dimensional set, parametrized by $\varphi $ above. What then
happens to the other
region determined by the first gauge condition?
The answer is that, at exactly $t=\infty (\sigma gauge)$, the
condition
$a+b=0$ suffices to describe all the regions, $i.e.$ at $t=\infty $
gauge orbits
of the two regions connect to yield a single gauge orbit, thus making
the
theory one dimensional.\\

Eq.(6) also allows an interesting description of the black hole
for finite but
large $t$. The point is that if the field $x$ is required to vary
rapidly
compared with the Liouville field $\varphi $, then the last two
terms in(6)
can be ignored compared with the first three; and one obtains a
Liouville
field $\varphi $ coupled to a $C=1$ matter field $x$. This novel
connection
between two dimensional gravity and two dimensional black hole
theories $i.e.$
the former theory being the limit of a large boost on the
$U(1)$ gauge group
of the latter, for quasistatic Liouville background, may
provide a means of
a detailed and explicit comparison of various aspects of the two
theories, in
particular their spectrum. To this end and as a first step, the
vertex operators
of the two theories will be related in the following.\\

Vertex operators of the $2d$ black hole theory can be written in
terms of the
matrix elements of a group element $g(u,v)$ between states labled
by the
Casimir operator $\overline {J}_0^2$ of $SL(2,R)$ and eigenvalues
of the third component
of the zero modes of the current,\\
\begin{equation}
\begin{array}{ll}
J^3_0:\lambda ,\omega >=\omega :\lambda ,\omega > & \\
\overline {J}_0^2 :\lambda ,\omega >=(-\frac {1}{2}+i\lambda )
:\lambda ,\omega >, &
\omega ,\lambda \ real.\end{array}
\end{equation}

It can be shown that the only nonzero matrix elements are [7],\\
\begin{equation}
V(u,v)=<\lambda ,\omega :g(u,v):\lambda ,-\omega >,
\end{equation}\\
which in the region I of Ref.1, and in terms of the parameters
$r$ and $\tau $ defined by\\
\begin{equation}
u=sh ^2 \frac{r}{2}e^{\frac {\tau }{2}}\ \ \ \ ,\ \ \ \
v=-sh^2 \frac{r}{2} e^{-\frac {\tau }{2}},
\end{equation}\\
become\\
\begin{equation}
V (r,z)=e^{i\tau \omega }<\lambda ,\omega :g(r):\lambda ,\omega >,
\end{equation}\\
where $g(r)$ is a Lorentz transformation generated by $\sigma _1$.\\
The matrix elements in (10) are Lagendre functions whose
asymptotic behaviour
leads to\\
\begin{equation}
V \sim e^{-\frac {r}{2}}[e^{-i(\lambda r+2\omega \tau )}+S
(\lambda ,\omega)
e^{i(\lambda r-2\omega \tau )}],
\end{equation}\\
with\\
\begin{equation}
S(\lambda )=\frac {\Gamma (1+i\lambda )\Gamma ^2 (\frac
{1}{2}-4i\lambda)}
{\Gamma (1-2i\lambda )\Gamma ^2(\frac{1}{2}-2i\lambda)},
\end{equation}\\
giving the amplitude for string scattering in the black hole
background [7].\\

Similarly, one can obtain the vertex operators and scattering
amplitude for the
$2d$ gravity theory by considering Liouville field to be an
$SL(2,R)/U(1)$
theory as discussed in $[5,6,7]$ and take the $C=1$ matter to
be some timelike
variable. In the case at hand these vertex operators may be obtained
asthe limit of the corresponding black hole vertex operators in (10)
with the boosted $U(1)$ sub group. The boosted vertex operators
will then be matrix
elements of a group element $g$ between states which are eigenvectors
of the
boosted $U(1)$,\\
\begin{equation}
V_t(u,v)=<\lambda ,\omega _t:g(u,v):\lambda ,-\omega _t>,
\end{equation}\\
\begin{equation}
:\lambda ,\omega _t>=e^{-\frac {t}{2}\sigma _1}:\lambda ,\omega>.
\end{equation}\\

In the region determined by the gauge condition $a+b=0$, as
$t\longrightarrow \infty$
we expect to get $2d$ gravity vertex operators from (13). In fact as\\
\begin{equation}
:\lambda ,\omega _t>\longrightarrow :\lambda ,\chi >,
\end{equation}\\
Where $:\lambda ,\chi >$ is the eigenvector of $\sigma $ with
eigenvalue
$\chi ,a$ simple calculation leads to these vertex operators,\\
\begin{equation}
V_t(u,v)\longrightarrow e^{-2i\chi e^{-\varphi}x}e^{-\varphi }
K_{i\lambda }
(2\chi e^{-\varphi }),
\end{equation}\\
Where $K_{i\lambda }$ is the Bessel's function of the third kind.

A few comments are in order now: First, the result (16) confirms the
indentification of $x$ with the $C=1$ matter field; second, $\chi $,
the limit of $\omega _t$,
replaces the cosmological constant in the conventional treatment
of Liouville theory [7]; finally, energy of the $2d$ gravity state in
$\chi e^{-\varphi }$ which confirms the quasi-static nature of the
Liouville field in this formulation.

The scattering amplitude can also be read off eq.(16) with the
result,\\
\begin{equation}
S=\chi ^{2i\lambda }\frac {\Gamma (1+i\lambda )}{\Gamma (1-i\lambda)},
\end{equation}\\
in agreement with the expected result for $2d$ gravity [7].
Further study of
the relation between $2d$ gravity and $2d$ black hole theory using
the map
defined in this work, in particular the relation between discrete
states of the
two theories, is in progress.\\

The author wishes to thank H.Arfaei and M. Alimohammadi for
collaboration at
the early stages of this investigation; also gratefully acknowledges
useful
discussions with L.O'Raifeartaigh and P. Sorba on the $SL(2,R)$
formulation of
Liouville theory.\\ \\
\begin{center}
{\large \sc References}
\end{center}
\begin{enumerate}
\item E. Witten, Phys. Rev. D {\bf 44} (1991)314.
\item J. Distler and P. Nelson, Nucl. Phys. B {\bf 374} (1992)123.
\item S. Chaudhuri and J. Lykken, FERMI-PUB-92/169-T, June 1992.
\item T. Eguchi, H. Kanno, and S. Yang, contribution to this workshop.
\item A. Alekseev and S. Shatashvili, Nucl. Phys. B {\bf 323}
(1989)719.
\item L. O'Raifeartaigh, P.Ruelle, and I. Tsutsui, Phys. Lett.
B {\bf 258} (1991)359.
\item R. Dijkgraaf, E. Verlinde, and H. Verlinde, Nucl. Phys.
B {\bf 371} (1992)269.
\item F. David, Mod. Phys. Lett. A3 (1988)1651; J. Distler and
H. Kawai, Nucl. Phys. B {\bf 321} (1989)509.
\end{enumerate}
\end{document}